# Explicit Calculations on Small Non-Equilibrium Driven Lattice Gas Models

Wannapong Triampo,[*] I Ming Tang and Jirasak Wong-Ekkabut

*Department of Physics, Faculty of Science, Mahidol University, Bangkok 10400, Thailand*
*Capability Building Unit in Nanoscience and Nanotechnology,*
*Faculty of Science, Mahidol University, Bangkok 10400, Thailand*

We have investigated the non-equilibrium nature of a lattice gas system consisting of a regular lattice of charged particles driven by an external electric field. For a big system, an exact solution cannot be obtained using a master equation approach since the many-particle system has too many degrees of freedom to allow for exact solutions. We have instead chosen to study small systems as a first step. The small systems will be composed of between two and four particles having two or three possible values of some parameters. Applying periodic boundary conditions and a hard-core or an exclusion-volume constraint and imposing conservation of particle numbers via Kawasaki-type dynamics (particle-hole exchange), we are able to calculate the exact solutions of the steady-state relative probability density function, $r_i$, associated with each configuration of the small system.

## I. INTRODUCTION

In nature, the equilibrium state is the exception. In both real physical and biological systems, non-equilibrium phenomena are more common [1–3]. Non-equilibrium states display overwhelmingly rich and complex behavior (both deterministic and stochastic), such as phase segregation and separation, pattern formation, self-assembly, turbulence, and chaos. When studying systems in thermal equilibrium from a statistical mechanics viewpoint, one utilizes the framework established by Gibbs [4], *i.e.*, first specify the microscopic Hamiltonian of the system and then express the time-independent or stationary distribution over the configuration space in terms of the Boltzmann factor. The observable averages are then calculated using these distributions. This has allowed equilibrium statistical mechanics to reach a rather mature status. In contrast, there is no sound foundation for studying non-equilibrium phenomena, so these phenomena are far less understood and are much more difficult to study. At the moment, there is no well-established systematic analytical recipe for calculating the averages over known ensembles. Instead, one has to rely on simulations and/or computational approaches or work with small systems.

Part of the difficulty with non-equilibrium systems is that the distributions associated with these systems are generally time dependent. The time evolution of such systems are governed by a master equation. One possible approach is to study systems which have reached a non-equilibrium steady state where the distribution is time dependent [5]. This state remains non-Hamiltonian with no equivalent of a Gibbs measure. To attack this kind of problem, one typically can start with a master equation of the form

$$\frac{\partial P(C,t)}{\partial t} = \sum_{[C']} \{W[C' \longrightarrow C]P(C',t) - W[C \longrightarrow C']P(C,t)\} \quad (1)$$

and look for steady-state solutions. In the above, $P(C,t)$ is the probability of finding the system in the configuration $C$ at time $t$ with a given initial condition. The dynamics is specified by the transition rate $[C' \longrightarrow C]$. It is quite clear that Eq. (1) is nothing but a balance equation. The first sum on the right-hand side represents the "gain" terms, in which all configurations from which $C$ could possibly originate are summed over. The second sum is the "loss" terms, which contains all the possible ways the system can leave $C$. Thus, the non-equilibrium steady state is fully described by $P^*[C] = P[C, t \to \infty]$ at the microscopic level. For the majority of such steady states, there is a non-vanishing uniform "current" in configuration space.

If one is to understand the non-equilibrium phenomena occurring in real world systems such as the relaxation of an initial ordered or disordered state to its final steady state, the non-linearity of the feedback of the collective non-equilibrium behavior in many particle systems, and the nature of the temporal and spatial scaling of the system, it is important that the simple systems used to model real systems have relevance to real systems and capture the essence of the physics involved. Studying these simple systems can then provide insight into the nature of the real systems.

[*]E-mail: wtriampo@yahoo.com

From a theoretical point of view, it is extremely difficult to analytically solve the master equation, Eq. (1). Instead, one applies computational methods or analytic methods containing approximations such as mean-field theory, renormalization group, an so on. Dealing with a very small, but analytically solvable system, allows one to learn something from the system, especially when used together with simulation and other approximate calculations. For example, in Ref. 6 Zia and Blum carried out some calculations on a $2 \times 3$ system by using a model similar to ours at finite temperature, but with zero field. Their results showed the hallmarks of non-equilibrium phenomena that violate the fluctuation-dissipation theorem. Evans and Hanley [7] used a relatively small system to locate the melting point of soft discs. While the calculations using small system may not lead to better analytic methods for real systems, they may provide ways to check possible new approaches to the more complicated systems.

Motivated by the above, we aim to point to what can be learned about some non-equilibrium systems by performing calculations on a small system. To do this, we will study systems of charged particles based on the driven lattice gas model of Katz *et al.* [8]. In the lattice gas model, each lattice site can be empty or be occupied by a single charged particle. In addition to this excluded volume constraint, the particle interacts with an external, uniform field $E$. We consider the infinite temperature limit, which means that no other interactions between particles other than the correlation via the excluded volume are taken into account in our dynamics. This model differs from the ordinary Ising model introduced by Lenz in 1925 [9], in that the particle can hop to neighboring unoccupied sites with a rate specified through a bias in the rate of hopping along an external uniform driving field. The particle-vacancy (hole) exchanges follow the (conserved) Kawasaki dynamics [10] with the Metropolis rates [11]:

$$[C \longrightarrow C'] = \min\{1, \exp[-qE\Delta y]\}, \quad (2)$$

where $q(= \pm 1$ or $0)$ is the charge of each particle, $E$ is the magnitude of the uniform external field (pointing downward), and $\Delta y$ is the change in the coordinate (parallel to the field) of the particle and is equal to 1 $(-1)$ if the jump is up (down). This transition rate is used to satisfy the detailed balance condition to guarantee that when the system is started from a unstable non-equilibrium configuration, it will reach steady state. Some examples of real systems which may be connected to this system (for a larger system size) are certain fast ionic conductors or solid electrolytes with two different ion species acting as charge carriers [12, 13] or water-in-oil microemulsions in which the small water droplets suspended in oil can carry electric charge [14,15]. These systems have been generalized and studied to gain insight into both the theoretical and the experimental aspects of systems, such as the dynamics of ordering in bulk systems, following a rapid temperature quench [16] and the dynamics of phase disordering after a rapid increase in temperature [17].

This paper is organized as follows: In Section II, we present the details of the different models case by case. We show how to formulate the non-equilibrium master equation for each case. In Section III, we solve the models analytically and discuss the key results. Finally, we summarize and present some comments and remarks in Section IV.

## II. MODEL AND FORMULATION

In this section, we show how to set up the master equation for each case. Let N be number of particles and $q_i$ be the charge of $i^{th}$ particles.

### 1. Case 1. $N = 2, q_1 = 1, q_2 = -1$

We consider here a $2 \times 3$ lattice system consisting of two oppositely charged ($q = \pm 1$) particles. Each lattice site can be described by one of two states: occupied or unoccupied. The electric field results in a biased hopping of the particles along one of the lattice directions. For a positively charged particle, the field favors jumps along its direction, suppresses jumps in the opposite direction, and is neutral to jumps in the transverse directions. For negatively charged particles, the opposite is true. Imposing periodic boundary conditions in both directions, we obtain translational invariance. Conservation of particle numbers and the hard-core constraints (multiple occupancy being forbidden) are also imposed.

Since there are two types of particles, we will have altogether $^6P_2 = 30$ possible configurations. If the translational invariance condition due to the periodic boundary conditions is applied, the 30 possible configurations fall into five groups. The members of each group can be obtained from each other by a translation. (See Fig. 1).

Next, we derive the master equations describing the dynamics of this system. This is done as follows:

We begin by drawing the diagrams of the probability flow where the gain is represented by incoming arrows and the loss is represented by outgoing arrows as shown in Fig. 2(a). In that figure, we show only the first configuration (i/1) of each group since the rest are obtained in the same manner.

We next use Eq. (2) to calculate the transition rate $[C' \longrightarrow C]$. This is illustrated in the diagram shown in Fig.2(b) for the 1/1 configuration. The details of the straightforward calculation are given below.

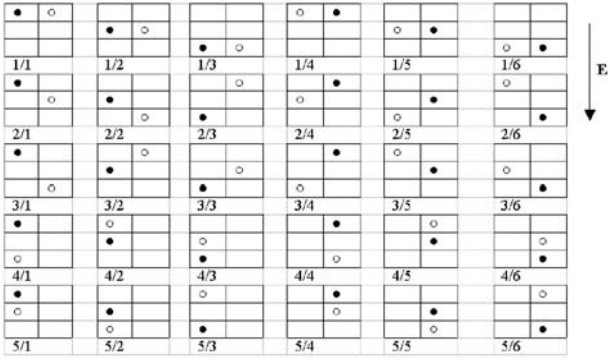

Fig. 1. All possible configurations of a 2 × 3 system for case1. (●) and (○) represent positively and negatively charged particles, respectively. The label i/j below each configuration indicates the $j^{th}$ configuration belonging to the $i^{th}$ group.

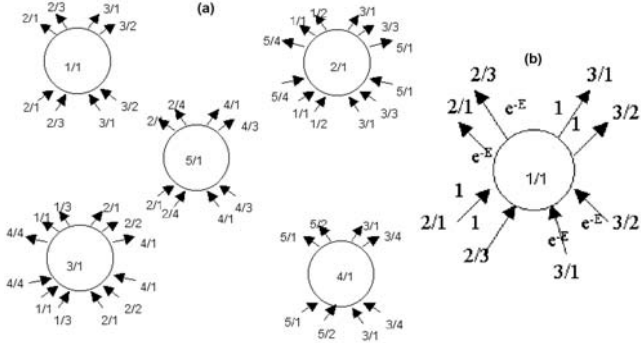

Fig. 2. Schematics of the configurational probability flow. The numbers inside the circles indicate the equivalent group as in Fig.1. The direction of the arrow represents direction of probability flow, the "gain" and "loss" terms

"Loss" contributions
$1/1 \longrightarrow 2/1$ the rate is min{1, exp[-(-1)E(-1)]} = $e^{-E}$
$1/1 \longrightarrow 2/3$ the rate is min{1, exp[-(+1)E(+1)]}= $e^{-E}$
$1/1 \longrightarrow 3/1$ the rate is min{1, exp[-(-1)E(+1)]}= 1
$1/1 \longrightarrow 3/2$ the rate is min{1, exp[-(+1)E(-1)]}= 1

"Gain" contributions
$2/1 \longrightarrow 1/1$ the rate is min{1, exp[-(-1)E(+1)]} = 1
$2/3 \longrightarrow 1/1$ the rate is min{1, exp[-(+1)E(-1)]} = 1
$3/1 \longrightarrow 1/1$ the rate is min{1, exp[-(-1)E(-1)]} = $e^{-E}$
$3/2 \longrightarrow 1/1$ the rate is min{1, exp[-(+1)E(+1)]} = $e^{-E}$

The same recipe can be used to calculate the other rates entering into $P_i$ ($i = 1, 5$).

We now write down the system of partial differential equations for the $P_i$'s defined by Eq. (1). These equations use the fact that many of the configurations are equivalent. The contributions from these configurations would have the same weights. Letting $x = e^{-E}$, we have

$$\partial_t P_1 = -(2 + 2x)P_1 + 2P_2 + 2xP_3$$
$$\partial_t P_2 = 2xP_1 - (4 + 2x)P_2 + 2P_3 + 2P_5$$
$$\partial_t P_3 = 2P_1 + 2xP_2 - (4 + 2x)P_3 + 2P_4$$
$$\partial_t P_4 = 2P_3 - 4P_4 + 2xP_5$$
$$\partial_t P_5 = 2P_2 + 2P_4 - (2 + 2x)P_5. \quad (3)$$

These equations can be written as a matrix equation $\partial_t P = wP$ where $\partial_t = \frac{\partial}{\partial t}$ and w is the matrix representation of the associated probability density matrix. The explicit matrix equation is

$$\begin{bmatrix} \partial_t P_1 \\ \partial_t P_2 \\ \partial_t P_3 \\ \partial_t P_4 \\ \partial_t P_5 \end{bmatrix} = \begin{bmatrix} -(2+2x) & 2 & 2x & 0 & 0 \\ 2x & -(4+2x) & 2 & 0 & 2 \\ 2 & 2x & -(4+2x) & 2 & 0 \\ 0 & 0 & 2 & -4 & 2x \\ 0 & 2 & 0 & 2 & -(2+2x) \end{bmatrix} \begin{bmatrix} P_1 \\ P_2 \\ P_3 \\ P_4 \\ P_5 \end{bmatrix} \quad (4)$$

This completes the formulation of the master equations. In the next section, we will focus on solving these equations.

**2. Case 2.** $N = 3, q_1 = q_2 = -1, q_3 = +1$

We now consider the same 2 × 3 lattice system, but with two negatively charged particles and one positively charged particle. Everything else is the same. Following the steps used in the previous case, we find that there are 60 ($\frac{^6P_3}{2!1!}$) configurations. The additional number is due to the presence of two indistinguishable particles. The 60 configurations can be classified into 10 groups (See Fig. 3).

Using the technique outlined in the previous case, we get the diagrammatic representation of the probability flux flow shown in Fig. 4.

We obtain the 10 × 10 probability densities matrix.

$$w = \begin{bmatrix}
-3 & 0 & 1 & 1 & 0 & 0 & 1 & 0 & 0 & 0 \\
0 & -(3+x) & x & x & 0 & 1 & 0 & 0 & 0 & x \\
1 & 1 & -(4+3x) & 1 & 1 & 1 & 1 & 0 & 0 & x \\
1 & 1 & 1 & -(6+x) & x & x & 1 & 0 & 0 & x \\
0 & 0 & x & 1 & -(2+3x) & 1 & 0 & 0 & 1 & 1 \\
0 & x & x & 1 & x & -(4+x) & 0 & 1 & 0 & 0 \\
1 & 0 & 1 & 1 & 0 & 0 & -(5+2x) & 1+x & 1+x & 0 \\
0 & 0 & 0 & 0 & 0 & 1 & 1+x & -(3+2x) & 1+x & 0 \\
0 & 0 & 0 & 0 & 1 & 0 & 1+x & 1+x & -(3+2x) & 0 \\
0 & 1 & 1 & 1 & x & 0 & 0 & 0 & 0 & -(3+x)
\end{bmatrix} \quad (5)$$

### 3. Case 3. $N=4, q_1=q_2=+1, q_3=q_4=-1$

We next consider the same lattice system, but with two negatively charged particles and two positively charged particles. For this system, we have 90 ($\frac{^6P_4}{2!2!}$) configurations. These 90 configurations can be classified into sixteen distinct groups (See Fig. 5).

The master equations for this system are obtained in the same manner as before. The resulting equations are given below:

$$ (6) $$

(16×16 matrix shown in Eq. 6)

### 4. Case 4. $N=3, q_1=1, q_2=0, q_3=-1$

We now consider the situation where there are three (instead of two as with the previous three cases) types of charged particles, negatively charged, positively charged, and a zero (neutral) charged. There is one of each on the lattice. Altogether, we will have 120 ($^6P_3$) configurations. Using the same recipe as before, we obtain the diagrams shown in Fig. 6.

The associated probability density matrix for this case is given by Eq. 7.

$$ (7) $$

(20×20 matrix shown in Eq. 7)

## III. EXACT RESULTS AND ANALYSIS

In this section, we show how the master equations derived in Section II can be solved. We will present in detail the analysis of case 1 only and quote the results for the other cases. Furthermore, we will only look at the steady-state case. Form Eq. (5) with $\partial_t P = 0$, we have

$$\begin{bmatrix} -(2+2x) & 2 & 2x & 0 & 0 \\ 2x & -(4+2x) & 2 & 0 & 2 \\ 2 & 2x & -(4+2x) & 2 & 0 \\ 0 & 0 & 2 & -4 & 2x \\ 0 & 2 & 0 & 2 & -(2+2x) \end{bmatrix} \begin{bmatrix} P_1^* \\ P_2^* \\ P_3^* \\ P_4^* \\ P_5^* \end{bmatrix} = 0 \qquad (8)$$

where $P_i^*$ represents the weight of a single configuration within group i at the steady state and not the weight for the group as a whole. Next, we row reduce the matrix to get

$$\begin{bmatrix} 1 & 0 & 0 & 0 & \frac{-x^3-4x^2-3x-3}{-3x^2-4x-4} \\ 0 & 1 & 0 & 0 & \frac{-x^3-3x^2-4x-3}{-3x^2-4x-4} \\ 0 & 0 & 1 & 0 & \frac{-x^3-4x^2-4x-2}{-3x^2-4x-4} \\ 0 & 0 & 0 & 1 & \frac{-2x^3-4x^2-4x-1}{-3x^2-4x-4} \\ 0 & 0 & 0 & 0 & 0 \end{bmatrix} \begin{bmatrix} P_1^* \\ P_2^* \\ P_3^* \\ P_4^* \\ P_5^* \end{bmatrix} = 0 \qquad (9)$$

The above form of the matrix equation clearly shows that $P_1^*$, $P_2^*$, $P_3^*$, and $P_4^*$ can be expressed in terms of $P_5^*$. The relative weights of these configurations at steady state are

$$r_1 = \frac{x^3 + 4x^2 + 3x + 3}{3x^2 + 4x + 4}$$

$$r_2 = \frac{x^3 + 3x^2 + 4x + 3}{3x^2 + 4x + 4}$$

$$r_3 = \frac{x^3 + 4x^2 + 4x + 2}{3x^2 + 4x + 4}$$

$$r_4 = \frac{2x^3 + 4x^2 + 4x + 1}{3x^2 + 4x + 4}. \qquad (10)$$

where $r_i = \frac{P_i^*}{P_5^*}$. When the external field $E$ is zero, $x = 1$ so $r_i = 1$. When the field is very large (i.e., $E \longrightarrow \infty$), $x = 0$, so $r_1 = r_2 = \frac{3}{4}$, $r_3 = \frac{1}{2}$, and $r_4 = \frac{1}{4}$. This is expected since the configuration in group 5 is the most probable one as can be explained by the following reason-ing. The positively charged particle energetically favors moving downward in the direction of the electric field while the negatively charged particle moves in the opposite direction. Given that a high field is present, there will only be a small probability for transverse jumps. This would result in there being obstruction between movement of the oppositely charged particle. Using the reverse reasoning, one can explain why group 4 is the least likely. One can see that the configurations in group 2 have only one transverse jump different from those in group 5. Subsequently, group 2 is the second most probable one. However, at very high field, $r_1 = r_2$ since one can ignore the transverse jump; i.e., only vertical jumps play a significant role.

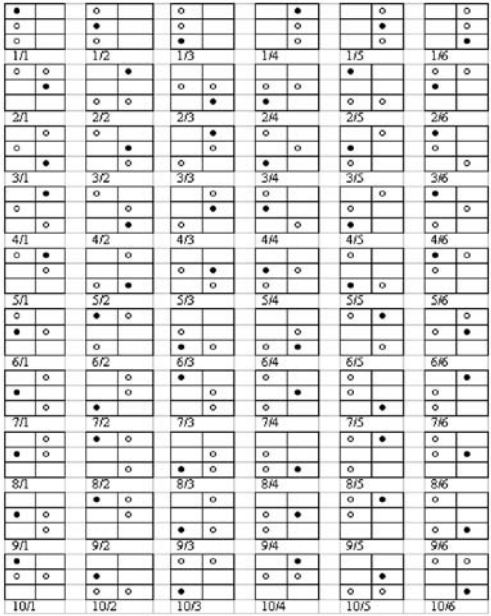

Fig. 3. All possible configurations of a 2 × 3 system for case 2. There are 1 positively and 2 negatively charged particles.

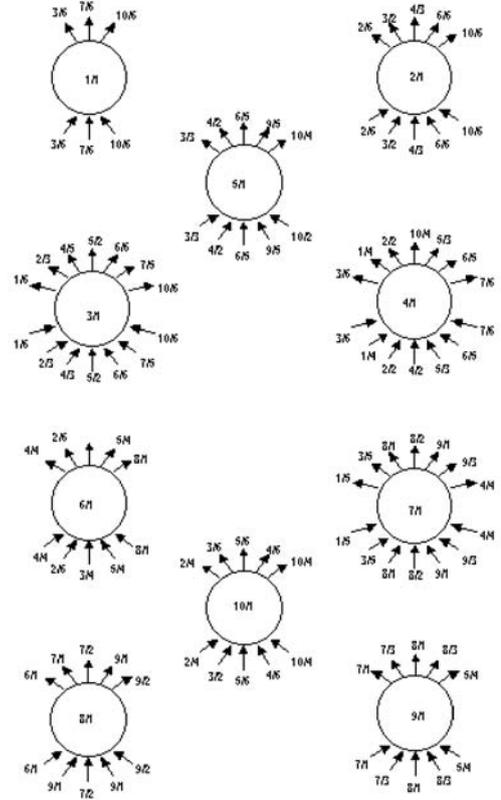

Fig. 4. Schematics of the configurational probability flow of the model in case 2. The notations are the same as those in Fig. 2.

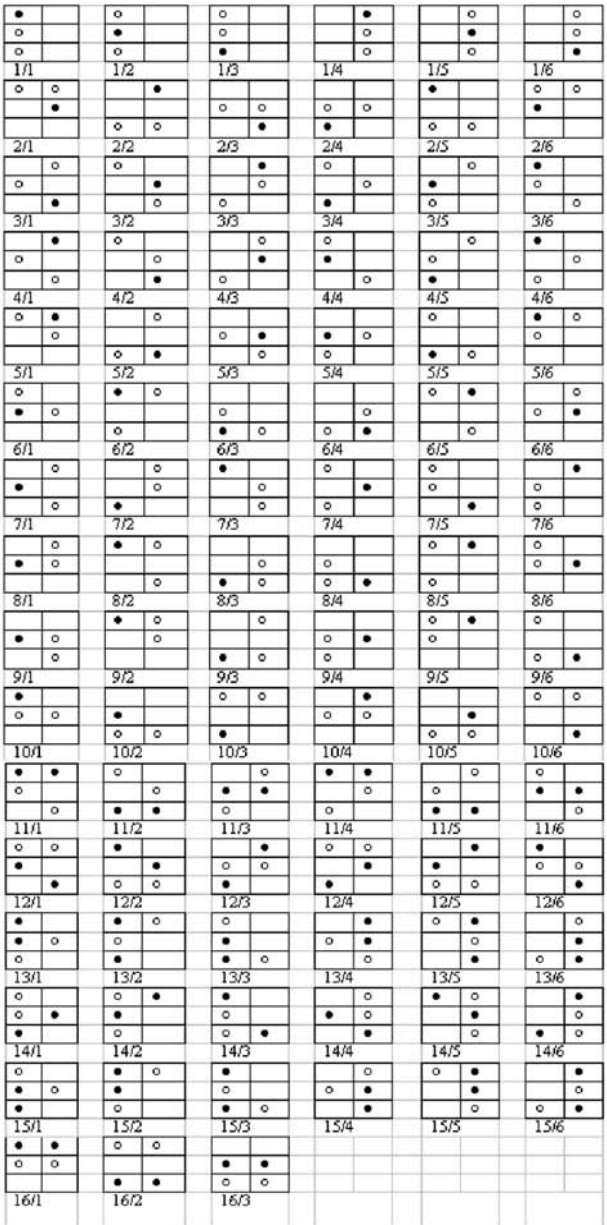

Fig. 5. All possible configurations of a 2 × 3 system for case 3. The notations used are the same as the previous ones.

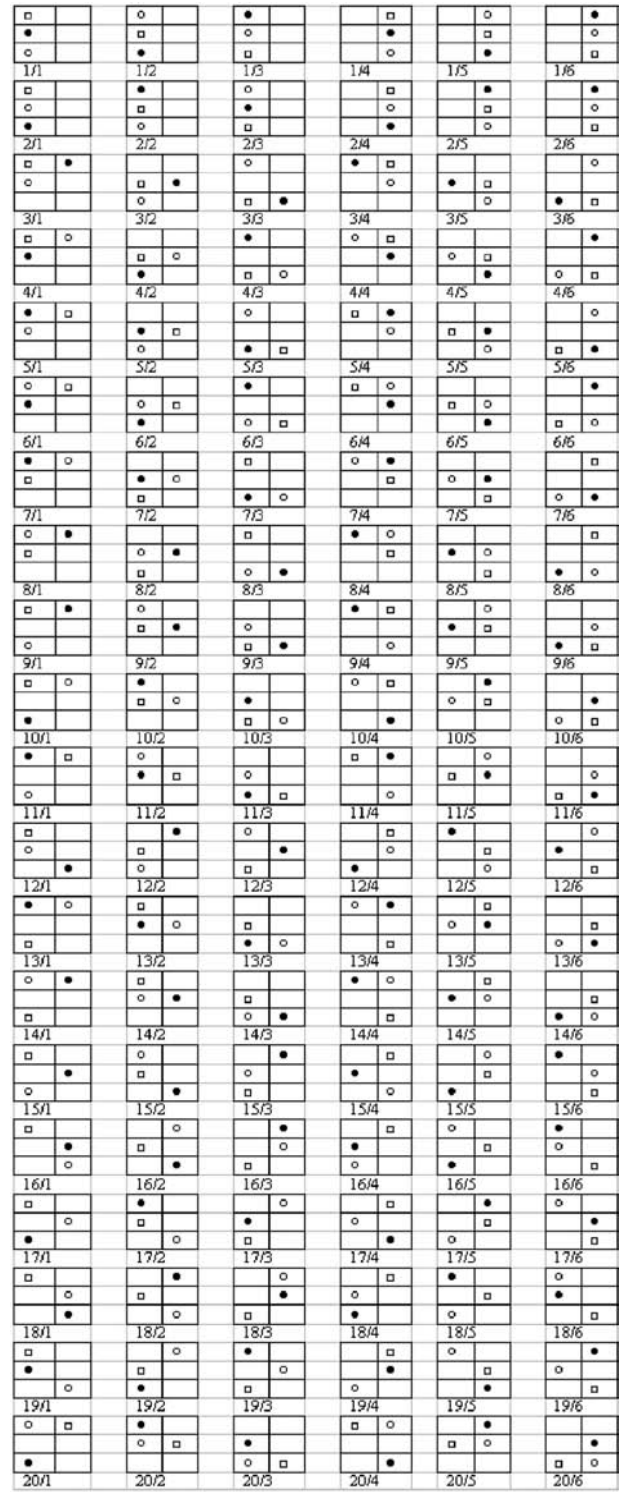

Fig. 6. All possible configurations of a 2 × 3 system for case 4.

Using similar methods, the relative weights of the configurations in cases 2 - 4 can be calculated. The results are shown in Figs. 8 - 10, respectively.

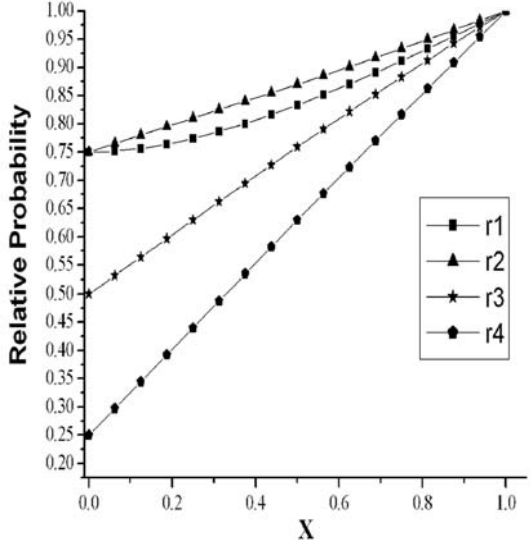

Fig. 7. Relative probability density function at steady state versus $x$ for case 1.

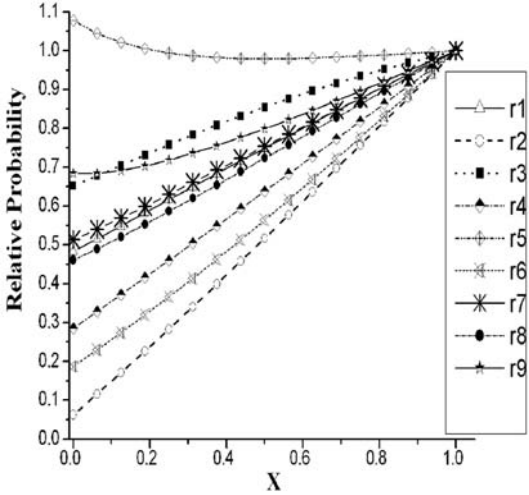

Fig. 8. Relative probability density function at steady state versus $x$ for case 2.

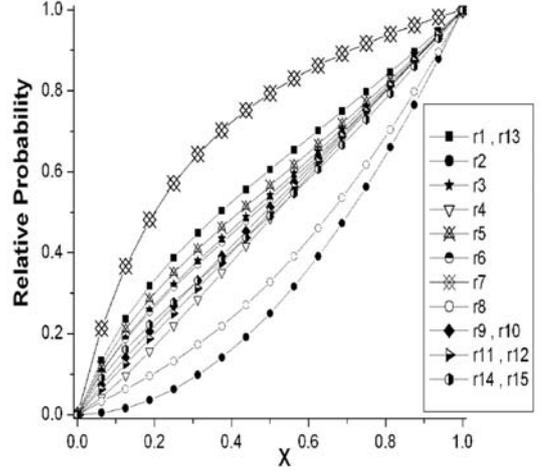

Fig. 9. Relative probability density function at steady state versus $x$ for case 3.

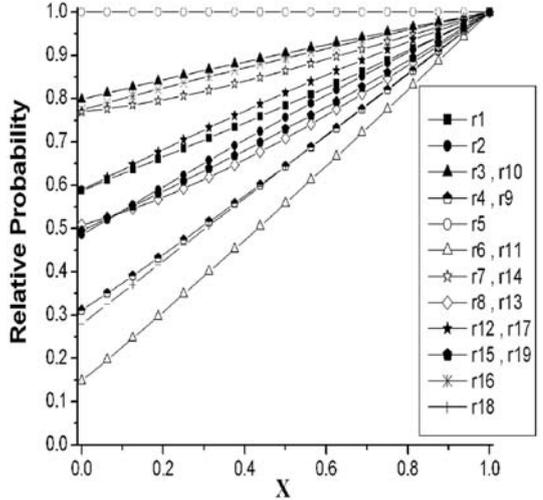

Fig. 10. Relative probability density function at steady state versus $x$ for case 4.

## IV. SUMMARY AND CONCLUSION

We have studied two- and three-species Driven Lattice Gas models on a $2 \times 3$ lattice system having fully periodic boundary conditions and an externally applied field through an exact calculation of the non-equilibrium master equations. The two-species system consists of positively charged particles and negatively charged particles. The three species system has, in addition, a neutrally charged particle. We have calculated the probability distributions $P_i^*$ at steady state and the relative probabilities for the equivalent groups. We have analyzed the steady-state solutions exactly.

For the two-species systems at steady state, we were able to find the relative probabilities, $r_i$, and find which configuration was most likely. Specifically, the configurations in groups 5, 10, and 16 are the most likely configurations for the two, three, and four particle systems, respectively. It should be noted that in the three-particle system, $P_5^*$ is bigger than $P_{10}^*$ for very large E and their ratio goes to one as $e^{-E}$ gets bigger. In the four-particle system, it is worthy to note that for very large E, the probabilities of all configurations go to zero, except for the configurations belonging to group 16. In these systems, it is not difficult to see why the configurations in group 16 are the most likely and those in group 7 are the second most likely, the reason being the tendencies of the positive particles to move downward (in the direction of the field) and of the negative particles to move upward. The results also show the "degeneracy" of the probabilities, namely $P_1^* = P_{13}^*$, $P_9^* = P_{10}^*$ and $P_{14}^* = P_{15}^*$. It is easy to see that the configurations in group 2 are the least likely.

For the three-species systems at steady state, we find the configurations in group 20 are the most likely. Because there are many more parameters appearing in this case, we were not able to obtain closed form expressions for the relative probabilities. One can see, however, that the graphs for this case are more or less, straighter (or more linear) than those for the two-species system.

We finish this section by mentioning how this work relates to a large system size and the connections between our systems and real systems. A specific example of a bigger size version which is closely related to our systems has been studied by Thies and Schmittmann [18]. In their paper, with the help of a Monte-Carlo simulation and a mean-field theory (coarse-grained level), they investigated the ordered steady-state structures resulting from the motion of a single vacancy on a lattice with periodic boundary condition, excluded volume constraint, and constant external field. The systems contained two species, positively and negatively charged, of particles and was initially disordered. They found that for a non-equilibrium steady state, the system underwent a charge segregation whose ordered steady-state configuration exhibited a phase separation approximately consistent with our prediction. Since we only wish to illustrate how to attack the non-equilibrium phenomena analytically by using exact method, we have neglected the effects of temperature. We have attempted to account for the violation of the fluctuation-dissipation theorem (a benchmark of non-equilibrium phenomena). Our calculation gives an indication on how the dynamic phase-segregated or ordered system occurs when starting from a homogeneous disordered system. However, to what extent the small system reflects what would happen in the large system when the number of species and/or density or particles are increased depends on the details of the dynamics one would apply.

## ACKNOWLEDGMENTS

The authors thank Drs. Michael Allen and Julian Poulter for reading the manuscript and providing helpful comments. We also thank an anonymous referee for his or her helpful comments and suggestions. This research is supported by a grant for the Thai Research Fund (TRF) through grants TRG 4580090 and RTA 4580005.